\documentclass[aps,prd,twocolumn,showpacs,groupedaddress,nofootinbib]{revtex4}
\usepackage{graphicx}
\begin{document}

\title{Sizes of the Lightest Glueballs in {\it SU(3)} Lattice Gauge
Theory}
\author{Mushtaq Loan}
\affiliation{Department of
Mechanics and Engineering, Zhongshan University, Guangzhou 510275,
China\\
School of Physics, The University of New South Wales, Sydney, NSW
2052, Australia}

\author{Yi Ying}
\affiliation{Department of Physics, Chong Qing University, Chong
Qing 400030, China}

\date{February 9, 2006}
\begin{abstract}
Standard Monte Carlo simulations have been performed  on improved
lattices to determine the wave functions and the sizes of the
scalar and tensor glueballs at four lattice spacings in the range
$a=0.05 - 0.145$ fm. Systematic errors introduced by the
discretization and the finite volume are studied. Our results in
the continuum limit show that the tensor glueball is approximately
two times as large as the scalar glueball.
\end{abstract}

\maketitle

\section{Introduction}

Lattice calculations have yielded very accurate spectroscopic
information regarding  low-lying hadrons \cite{UKQCD00,PACS00} and
glueballs
 \cite{Morningstar99,Vaccarino99,Teper98,Norman98,Bali93,Chris90}.
These calculations show that the lowest-lying scalar, tensor and
axial vector glueballs lie in the mass region 1 - 2.5 GeV. While
there is a long history of hadron and glueball mass calculations
in lattice QCD, progress in determining their wave functions has
not been so rapid. Although the spectrum of pure gauge QCD cannot
be directly compared to experiment, it is important to pin it down
because so little is known about glueball properties. The mixing
of glueballs with $q\bar{q}$ mesons due to the presence of  light
dynamical quarks makes it difficult in distinguishing them from
ordinary mesons, both in full lattice QCD and in  experiments
\cite{Sexton95}. Enormous effort have been made in pursuit of
glueballs and their properties \cite{Seth00}. Accurate lattice
calculations of their size, matrix elements and form factors would
help considerably in the attempt to realize their experimental
identification.

Glueball wave functions and sizes have been studied previously
 \cite{Forcrand92,DeGrand87,Ishikawa83, Ishii02}, but much of the
early works contain uncontrolled systematic errors, most notably
from discretization effects. However, calculations
using operator overlaps obtained from variational optimization for
an improved lattice gauge action showed that the scalar and tensor
glueballs are of typical hadronic dimensions. A straightforward
procedure to determine glueball size is to compute
measure the glueball wave function, much in the same way as in the meson
and baryon wave functions were measured.

Although Ref. \cite{Forcrand92} produced interesting results, the
approach used there suffers from a basic problem: The observables
are calculated from lattice version of the 2-glue operator, which
risks the mixture of glueball states with the flux
state\footnote{The link-link operator used  there sums up a large
number of loops; some of these loops have a zero winding numbers
and project onto glueballs others have non-zero winding numbers
and project onto flux states, also called torelons.}. Also the
results are of limited interest, because of their manifest
dependence on the gauge chosen. In this study we take a more
direct approach to the problem: Instead of fixing a gauge or a
path for the gluons, we measure the correlation functions from
spatially connected Wilson loops which, being the expectation
values of closed-loop paths, are gauge invariant. The procedure
has been used in the measurement of the wave functions of the
scalar and tensor glueballs in the lattice Monte Carlo
calculations in 3-dimensional {\it U(1)} theory \cite{Mushe06}.
This approach has the advantage that the detangling of the
glueball and torelon is usually taken care of automatically by the
choice of Wilson or Polyakov loops. This means that glueball
operators cannot create a single torelon state, but the creation
of torelon-antitorelon pairs is possible.

In this paper, we demonstrate the efficiency of our method in
calculating {\it SU(3)} glueball wave functions using the
tadpole-improved Symanzik action \cite{Norman98}. Since we want to
explore the nature of the wave function, we focus on the sizes and
masses of two of the lighter {\it SU(3)} glueball states, the
scalar and the tensor. To avoid  mixing effects, we adopt a
quenched lattice QCD as a necessary first step before attempting
to include the effect of dynamical quarks. It is worth mentioning
here that quenched lattice QCD accurately predicts the important
nonperturbative quantities as well as the masses of hadrons,
mesons, and baryons. The remaining content of this paper are
organized as follows. In Sec. \ref{sec2} we give a prescription
for calculating the glueball wave function  and size based on the
information from glueball correlators with the smearing method. We
present and discuss our results in Sec. \ref{sec3}. We close
section with a comparison of our results with related QCD results.
Concluding remarks are given in  Sec. \ref{sec4}.

\section{Wave functions of glueballs}
\label{sec2}
The lattice observables are computed as follows.
First, we calculate the lattice operator
\begin{equation}
\Phi_{i} (\vec{r},t)=\sum_{\bf x}\left[\phi_{i}(\vec{x},t)+
\phi_{i}(\vec{x}+\vec{r},t)\right],
\label{eqn01}
\end{equation}
where $\phi$ is the plaquette operator and $\Phi$ is the
two-plaquette or two-loop component of the glueball wave function.
The summation over $x$ is performed in order to obtain  the zero-momentum
projection. The $r$ dependence is reflected in the length of the
links required to close the loops.

Given a pair of orbit and spin quantum numbers $L$ and $S$ for
the desired glueball, one can construct operators
$W(\mid\vec{r}\mid ,t)$ with $J=0,1,2$, $P=\pm$, and $C=\pm$ from
suitable linear combinations of the rotation, parity inversions and
real and imaginary parts of the operators involved in $\Phi$. In
this study, we only consider $S=L=0$ (scalar $(0^{++})$ and $S=2$, $L=0$
(tensor $2^{++}$) glueball states. The two
lattice observables measured are therefore
\begin{eqnarray}
W_{0}(\mid\vec{r}\mid,t) & = & \sum_{r}\left[\Phi_{12} +\Phi_{13}
+\Phi_{23}\right],\\
W_{2}(\mid\vec{r}\mid,t) & = & \sum_{r}\left[2\Phi_{12} -\Phi_{23}
-\Phi_{13}\right].
\label{eqn02}
\end{eqnarray}

Using the glueball operators $W$, we consider the correlator
\begin{equation}
C(\vec{r},t)=\langle W^{\dagger}(\vec{r},t)W(\vec{r},0)\rangle ,
\label{eqn02b}
\end{equation}
where the vacuum expectation value must to be subtracted for the $0^{++}$
glueball. The source can be held fixed, while the sink takes on the $r$
dependence. This proves to be helpful in maintaining a good
signal. In general, to measure the ground-state mass from the
correlator, one seeks  the region where the overlap with the lowest
state is maximum and contributions from excited states are negligible.
In principle, such a region always exists for large Euclidean
time $t$ in the zero-temperature case. However, in practice, it is
difficult to use such large $t$ in lattice QCD Monte Carlo
calculations, since the correlator decreases exponentially with $t$
and becomes so small for large $t$ that it is comparable to its
statistical errors. Hence, for the ground-state mass measurement,
it is important that the ground-state overlap  be
sufficiently large. This leads to what amounts to a discrete search among wave
functions. In the case of glueballs in quenched {\it SU(3)} lattice QCD,
the overlap of the operators given in Eqs. (2) and (3) is quite
small as long as they are constructed from the simple plaquette
operators. This small overlap originates from the fact that the ``size"
of the plaquette operator is  smaller (specifically of $O(a)$) than the
physical size of the glueball \cite{APE87}. To improve the
glueball operator, to enhance the ground-state contribution, we
exploit the APE link smearing techniques \cite{APE87,Takahashi01}
to generate extended operators with sizes that are approximately the same
as the physical size of the glueball. The smearing procedure
is implemented through the iterative replacement of the original
spatial link variable by  a smeared link. For the smearing
parameters, which play an important role in extracting the
ground-state contribution, we used a smearing fraction of $\alpha
=0.5$ and a smearing number $N_{\rm{smr}}=15$ in the present
calculation. The optimum smearing is determined by examining the
ratio
\begin{displaymath}
C(r,t+1)/C(r,t)
\end{displaymath}
which should be close to unity for good ground sate dominance. In
fact, in the range  $4\leq N_{\rm{smr}}\leq 11$, the above
ratio is almost unchanged. For our measurements, with $\alpha
=0.5$ fixed, a typical value that proved  to be sufficient for
most cases was $N_{\rm{smr}}=4$ (although the results obtained with this value
are almost the same as those with $N_{\mbox{smr}}=11$). For fixed $\alpha$,
$N_{\rm{smr}}$ plays the role of extending the size of the
smeared operator and hence can give the rough estimate of the
physical size of the glueball.

A second pass was made to compute the optimized correlation
matrices
\begin{equation}
C_{ij}(t)=\langle W (r_{i},t)W (r_{j},0)\rangle - \langle
W(r_{i})\rangle \langle W (r_{j})\rangle .
\label{eqn03}
\end{equation}
Let $\psi^{(k)}$ be the radial wave function of the $k$-th
eigenstate of the transfer matrix. Then we have
\begin{equation}
C_{ij}(t)=\sum_{k}\alpha_{k}\psi^{(k)}(r_{i})\psi^{(k)}(r_{j})e^{-m_{k}t}.
\label{eqn04}
\end{equation}
The glueball masses and the wave functions are extracted from the
Monte Carlo average of $C_{ij}(t)$ by diagonalizing the
correlation matrices $C(t)$ for successive times $t$,
\begin{equation}
C(t)=\tilde{R}(t)D(t)R(t),
\label{eqn05}
\end{equation}
where $D$ is a diagonal matrix whose elements are eigenvalues and $R$ is a
rotation matrix whose columns are the eigenvectors of $C$. Each
eigenvector of $C$ corresponds to an eigenstate $\psi^{(k)}(r)$ of the
complete transfer matrix.

In order to  investigate the size of the glueball, the
Bethe-Salpeter amplitude provides a convenient tool. In the
nonrelativistic limit, it is expected to reduce to the glueball
wave function in the first-quantized picture. This property can be
exploited to estimate the size of the glueball \cite{Forcrand92}.
Similar to the case for mesons, the wave function is expected to
decrease exponentially with the separation $r$ and is therefore
fitted with the simple form
\begin{equation}
\psi (r) \equiv e^{-r/r_{0}}
\label{eqn06}
\end{equation}
to determine the effective radius $r_{0}$.

The signal in the connected 2-point correlator $C(t)$  for all
observables does not last long to make to an exponential fit.
For this reason, the  data are  presented in terms of the effective
mass  read  directly from the largest eigenvalue corresponding
to the lowest energy,
\begin{equation}
m_{\mbox{eff}} =
\mbox{log}\left[\frac{\lambda_{0}(r=0,t=1)}{\lambda_{0}(r=0,t=2)}
\right].
 \label{eqn07}
\end{equation}
However, to ensure the validity of our results, we compared them to those obtained
 using
\begin{equation}
m^{'}_{\mbox{eff}} =\left[\frac{\lambda_{0}(t-1)-\lambda_{0}(t)}{\lambda_{0}(t)-\lambda_{0}(t+1)}\right].
\label{eqn07b}
\end{equation}
\section{Simulations results and discussion}
\label{sec3}

An ensemble of gauge configurations were generated using a combination
of the Cabibbo-Marinari (CM) algorithm and the overrelaxed method.
Configurations were given a hot start and then 100 compound sweeps
(we define one compound sweep as one CM update followed by five
over-relaxation sweeps) in order to equilibrate. After
thermalization, configurations were stored every 50 compound sweeps
for 750 configurations. Measurements made on the stored
configurations were binned into 10 blocks with each block
containing an average of 75 measurements. The mean and the final
errors were obtained using the single-elimination jackknife method with
each bin regarded as an independent data point. Four sets of
measurements were taken to check the scaling of our results at
$\beta = 2.0, 2.25,2.5$ and 2.75 on $16^{3}\times 16$ lattice.
Some finite-size consistency checks were done  on a $12^{3}\times
12$ lattice. Even though this lattice is relatively coarse, we can safely
conclude that there are no large discretization errors after the
implementation of the improved gauge and fermion actions.

\begin{figure}[!h]
\scalebox{0.45}{\includegraphics{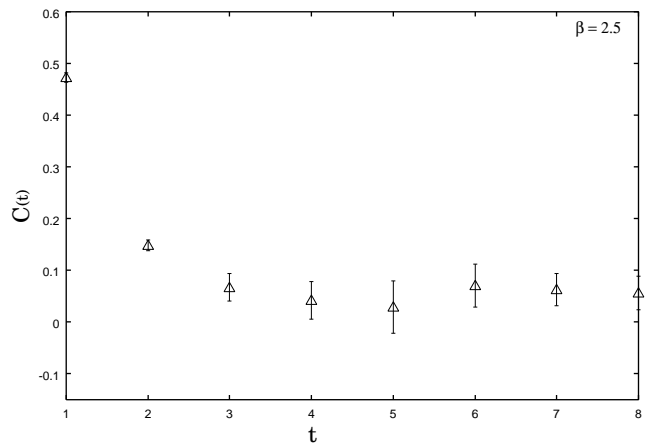}} \caption{
\label{figcorr} Correlation function for the scalar glueball at
$\beta =2.5$ on a $16^{4}$ lattice}
\end{figure}

The glueball correlation function for the $0^{++}$ channel as a
function of $t$ at $\beta =2.5$ is shown in Fig. \ref{figcorr}.
The correlation function exhibits the expected exponential
behaviour with respect to Euclidean time $t$ for small values of
$r$. Concerning the determination of the tensor glueball, we
observe that the signal for the $E^{++}$ channel is better than
that for the $T_{2}^{++}$ channel, and errors on the effective
mass are reasonable. For this reason, we choose to use $E$ of the
tensor at finite lattice spacing.  At $\beta=2.0$, the signal does
not persist long enough to demonstrate convergence to the
asymptotic value; we present results for time separations $t =1$
and 2. We found that Eqs. (\ref{eqn07}) and (\ref{eqn07b}) yield
consistent results in all cases analyzed here. The best estimates
for the masses from our long runs are collected in Table
\ref{tab1}.

\begin{figure}[!h]
\scalebox{0.45}{\includegraphics{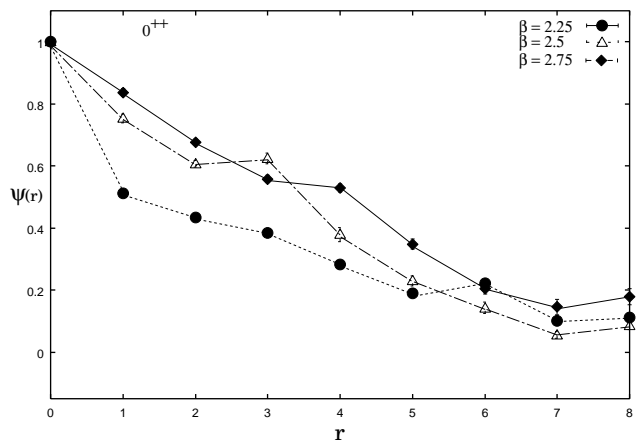}}
 \caption{ \label{fig0pp}
{\it SU(3)} scalar glueball wave functions computed on a $16^{4}$
lattice for various values of $\beta$.}
\end{figure}

\begin{figure}[!h]
\scalebox{0.45}{\includegraphics{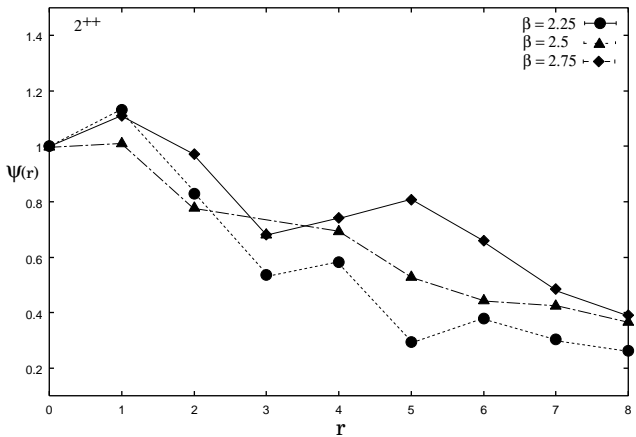}}
\caption{ \label{fig2pp}
Wave functions of the tensor glueball computed on a $16^{4}$
lattice with $\beta =2.25,2.5$ and 2.75.}
\end{figure}

We extracted the wave functions at time-separations $t=1$. Little
difference between the eigenvectors of $C(t)$ at $t=1$ and 2 was
found. This  suggests that there is no mixing between states of
distinct masses. Typical plots of the wave functions, normalized
to be unity at the origin, for the scalar and tensor glueballs at
$\beta = 2.25, 2.5$ and 2.75 are shown in Figs. \ref{fig0pp} and
\ref{fig2pp}, respectively. As a guide to the eye, Monte Carlo
points corresponding to of the same value of $\beta$ are connected
with straight lines. The scalar wave function exhibits the
expected behaviour and remains positive for all the values of
$\beta$  analyzed here.

We find that the tensor wave function shows the expected flatness
and is much more extended than the scalar one as  $\beta$
increases. This would imply that tensor is  more sensitive to
finite-size effects, which is very visible in the distortion of
the wave function for large $r$ at $\beta = 2.75$. Naively, we
expect that the spatial size at which we begin to encounter large
finite-size effects is related to the size of the glueball. From
optimization analysis at $t=1/0$, we found that the $4\times 4$
and $5\times 5$ loops have  better overlap with the glueball. This
suggests that the glueballs have a size of $\sim 4a - 5a$. The
smaller loops exhibit a weak signal and a slow convergence with
$N_{\rm{smr}}$. This is consistent with the findings of APE in the
case that a  $1\times 1$ loop is used as a template \cite{APE87}.

\begin{figure}[!h]
\scalebox{0.45}{\includegraphics{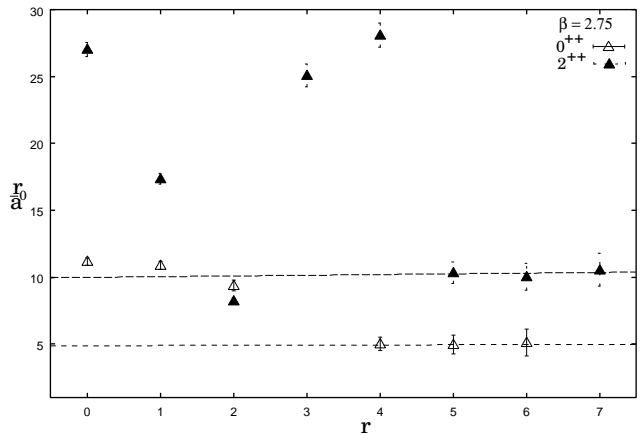}}
\caption{
\label{figeffrad} The effective radius  for the scalar and tensor
glueball states at $\beta=2.75$. The dashed horizontal lines
indicate the plateau values.}
\end{figure}

The size of the glueball is actually a nontrivial quantity.
Although the charge radius of the glueball can be formally
defined, its electric charges are carried by the quarks and
antiquarks, which play only  secondary roles in describing the
glueball state in the idealized limit, since the glueball does not
contain any valence content of quarks (antiquarks). In principle,
one could use Eq. (\ref{eqn06}) to extract the effective radius,
but because of the distortion of the wave function at large $r$
the results would depend strongly on the fit analysis of the above
form. Even after the ground-state enhancement, the complete
elimination of all the contributions of the excited states is
impossible, especially near $r \simeq 0$. It follows that Eq.
(\ref{eqn06}) holds only on a limited interval, which does not
include the vicinity of $r\simeq 0$. Hence, for the accurate
measurement of the effective radius, we need to find an
appropriate fit range. To this end, we examine a plot of the
effective-radius  and the ratio
\begin{displaymath}
\mbox{log}\left[\frac{\psi (r)}{\psi(r+1)}\right]
\end{displaymath}
for a given $\psi (r)/\psi(r+1)$ at each fixed $r$. In Fig.
\ref{figeffrad} we plot the effective radius $r_{0}$ as a function
of  $r$ associated with Figs. \ref{fig0pp} and \ref{fig2pp} at
$\beta=2.75$. Figure \ref{figeffrad} shows a  nontrivial $r$
dependence in the neighborhood of $r\simeq 0$  in the effective
radius of the tensor state. However, at large $r$, we see that
there appears a region where $r_{0}$ takes almost a constant
value. Thus we conjecture  that $r_{0}$ consists of nearly a
single spectral component, and hence can properly represent the
effective radius. Estimates of the sizes, in lattice units, at
various values of $\beta$ values are listed in Table \ref{tab2}.

In order to ascertain the finite-size effect on our measurements,
we performed extra simulations for $\beta = 2.5 $ and 2.75, using
a lattice of spatial extent $L= 12$. The results from this spatial
extension for glueball masses in terms of lattice units are given
in Tables \ref{tab1} and \ref{tab2}, respectively. Note that the
results for the smaller volume differ very little from those for
$16^{3}$ lattice, indicating that systematic errors in these
results attributable to the finite volume are negligible. We also
find that our estimates for the  size of glueballs are consistent
with the assumption that there is no large finite volume
dependence at $\beta=2.75$. Given that we use the larger lattice
size to extract our results we can disregard any contamination of
the glueball signal by a torelon-antitorelon pair, since no mass
reduction of sufficient magnitude was found as the lattice volume
was reduced (see Table \ref{tab1}). This suggests that none of our
states could be interpreted as a torelon pair.
\begin{table}[!h]
\caption{ \label{tab1} Scalar  and tensor glueball energy
estimates in lattice units for  the spatial extensions $L=12$ and
$L=16$. Also  shown are our estimates of the rho mass at a
physical point on the $L=16$ lattice.}
\begin{ruledtabular}
\begin{tabular}{ccccccc}
   & &  \multicolumn{3}{c}{Mass}\\
    & \multicolumn{2}{c}{$am_{0^{++}}$}& \multicolumn{2}{c}{$am_{2^{++}}$}
& $am_{\rho}(\kappa_{c})$ \\
 $\beta$/L & 12 & 16 & 12  & 16  & 16 \\ \hline
2.0  &  & 0.801(5) &  & 1.42(2) & 0.563(3) \\
2.25 &  & 0.58(1)&  & 1.07(2)  & 0.391(4)\\
2.5  & 0.52(2) & 0.527(3) & 0.82(2)  & 0.824(16) & 0.288(4) \\
2.75 & 0.43(2) & 0.443(4)& 0.64(2) & 0.659(14) & 0.227(4)\\
\end{tabular}
\end{ruledtabular}
\end{table}

\begin{table}[!h]
\caption{ \label{tab2} Effective radii of scalar and tensor
glueballs in lattice units for the two spatial extensions $L=12$
and $L=16$.}
\begin{ruledtabular}
\begin{tabular}{cccccc}
   & &   \multicolumn{2}{c}{Size}\\
   & \multicolumn{2}{c}{$r_{0}^{0^{++}}/a$}&
\multicolumn{2}{c}{$r_{0}^{2^{++}}/a$} \\
 $\beta$/L & 12 & 16 &12  & 16   \\ \hline
2.0   &  & 1.37(7)  &  & 3.08(14) \\
2.25  &  & 2.9(1) &  & 5.5(5)\\
2.5   & 4.1(1)  & 4.23(4)  &7.6(8)  & 7.5(6)\\
2.75  &4.8(2)   & 4.77(6)& 10.0(1.3) & 9.95(1.24)
\end{tabular}
\end{ruledtabular}
\end{table}
To convert scalar and tensor glueball sizes and masses to physical
units and extrapolate to the continuum limit, we first need an
estimate of the lattice spacing. One natural choice for this
conversion factor is the rho mass, $m_{\rho}a$. Estimating the
lattice spacing can be done by extrapolating the $\rho$ mass to
the physical quark mass\footnote{ Vaccarino and Weingarten
\protect\cite{Vaccarino99} noted that extrapolating to zero
lattice spacing using $[\Lambda\frac{(N_{f}=0)}{MS^{a}}]$ should
give results nearly equivalent to those found using
$m_{\rho}(a)a$.}. Using a tadpole-improved clover fermion action
we obtain the estimate $1/\kappa_{c}=6.362(3)$ by extrapolating
the  square pion masses from the largest five $\kappa$ values at
$\beta = 2.5$. We find that a linear fit of $1/\kappa$ to
$m^{2}_{\pi}$ in $1/\kappa$ works well for the rest of the
coupling values analyzed here. Linearly extrapolating the $\rho$
mass to $\kappa_{c}$, we find $am_{\rho}(\kappa_{c})=0.288(4)$ or
$a^{-1}=2.66(5)$ GeV, where the error is a jackknife estimate. The
lattice spacings at other values of $\beta$ are listed in Table
\ref{tabcont}.

Another type of error that might effect the simulation results
comes from the scaling violation for our action. Expecting  that
the dominant part of the scaling violation errors from the gauge
and light quark sectors are largely eliminated by the tadpole
improvement, we extrapolated the results obtained with finite $a$
to the continuum limit, $a\rightarrow 0$. In practice, it is often
difficult to quantify the magnitude of the systematic errors
arising from this origin. Here, we adopt an $a^{2}$- linear
extrapolation for the continuum limit, because the leading-order
scaling violation is always $O(a_{s}^{2}\Lambda_{\mbox{\rm
QCD}}m_{q})$. We also performed an $a$-linear extrapolation to
estimate systematic errors. We use the results obtained with the
three finest lattice spacings for the continuum extrapolation,
excluding the results obtained at $\beta = 2.0$, which appear to
have large discretization errors as expected from the naive order
estimate.

\begin{table}[!h]
\caption{ \label{tabcont} The lightest {\it SU(3)} glueball energy
and size  estimates in terms of the $\rho$ mass.}
\begin{ruledtabular}
\begin{tabular}{cccccc}
$\beta$ & $a(\mbox{fm})$ & $m_{0^{++}}/m_{\rho}$ &
$m_{2^{++}}/m_{\rho}$ & $r_{0}^{0^{++}}m_{\rho}$ &
$r_{0}^{2^{++}}m_{\rho}$ \\ \hline
2.0  & 0.1444(6) & 1.422(8)& 2.53(4) &  0.77(5) & 1.73(4) \\
2.25 & 0.100(2)  & 1.50(2) & 2.76(6) &  1.15(4) & 2.17(9)\\
2.5  & 0.073(3)  & 1.83(3) & 2.86(4) &  1.18(3) & 2.2(1)\\
2.75 & 0.058(6)  & 1.95(4) & 2.87(5) &  1.12(5) & 2.26(14)\\
\end{tabular}
\end{ruledtabular}
\end{table}
Performing such extrapolations for mass and size we adopt the
choice that yields the smoothest scaling behaviour for the final
value, and use others to estimate the systematic errors.

\begin{figure}[!h]
\scalebox{0.45}{\includegraphics{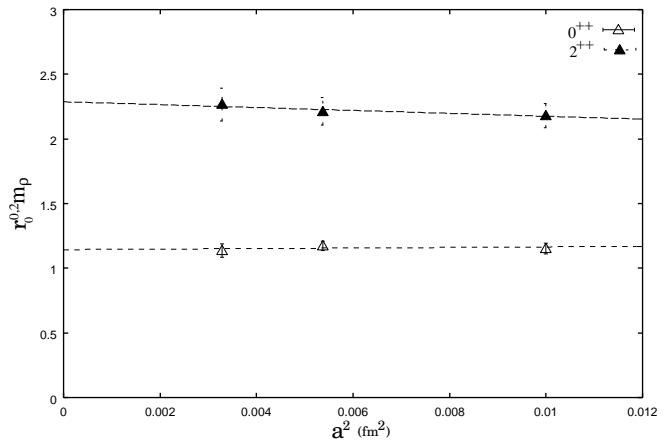}}
\caption{\label{figconsize}
Continuum limit extrapolation of glue radii in terms of $\rho$ mass.  The
dashed curves are the best fits to the simulation results over the
range $0.0033\leq a^{2}\leq 0.01$.}
\end{figure}

Figure \ref{figconsize} displays our results for the scalar and
tensor glueball radii in terms of $m_{\rho}$. It is seen that
linear extrapolations in $a^{2}$ yield  better fits and the value
in the continuum limit of $1.14\pm 0.08$ and $2.27\pm 0.03$ for
the scalar and tensor states, respectively.

\begin{figure}[!h]
\scalebox{0.45}{\includegraphics{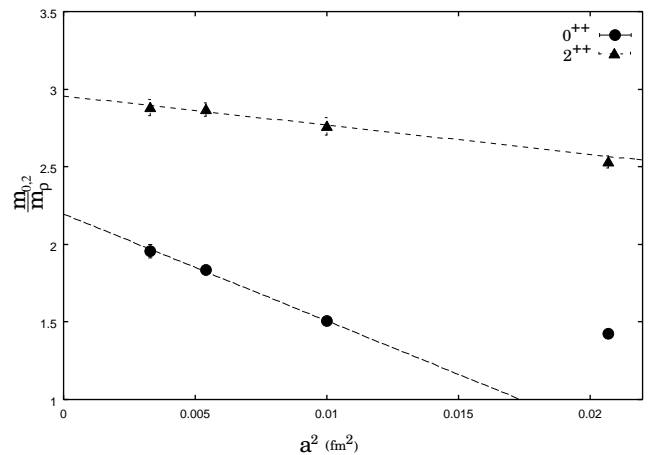}}
\caption{\label{cmass}
Continuum limit extrapolation of the glue energy estimates in
terms of the $\rho$ mass. The curves represent linear fits to the
data over the range $0.0033 \leq a^{2}\leq 0.01$.}
\end{figure}

Note that the product $r_{0}^{0,2}m_{\rho}$  varies only slightly
over the fitting range. The three non-zero lattice spacing values
of the product are within 0.02 - 0.04  and 0.01 - 0.09 standard
deviations of the extrapolated zero lattice spacing result for the
scalar and tensor state, respectively. This allows unambiguous and
accurate continuum extrapolations. Figure \ref{cmass} plots the
dimensionless ratio $m_{0,2}/m_{\rho}$ as a function of the square
of the lattice spacing. Again, we expect a linear fit in $a^{2}$
to provide the most reliable extrapolation to the $a\rightarrow 0$
limit. Linear extrapolations to the continuum limit yield a scalar
estimate of $2.19\pm 0.06$ and a tensor result of $2.95\pm 0.12$.
In contrast to the tensor, the scalar glueball mass shows
significant finite-spacing errors. The continuum limit results
obtained, in terms of the rho mass, are summarized in the Table
\ref{tabf}.
\begin{table}[!h]
\caption{ \label{tabf} Continuum limit predictions for the scalar
and tensor glueball masses and sizes and their conversion to MeV
and fm, respectively.}
\begin{ruledtabular}
\begin{tabular}{cc}
$m_{0^{++}}/m_{\rho}$ & $2.19\pm 0.06$ \\
$m_{2^{++}}/m_{\rho}$ & $2.95\pm 0.12$ \\
$r_{0}^{0^{++}}m_{\rho}$ & $1.14\pm 0.08$ \\
$r_{0}^{2^{++}}m_{\rho}$ & $2.27\pm 0.03$ \\
$m_{0^{++}}$ & $1680\pm 46$ MeV\\
$m_{2^{++}}$ & $2265\pm 92$ MeV\\
$r_{0}^{0^{++}}$ & $0.29\pm 0.02$ fm\\
$r_{0}^{2^{++}}$ & $0.58\pm 0.007$ fm
\end{tabular}
\end{ruledtabular}
\end{table}

To obtain the masses and the radii in units of MeV and fm, we used
the experimental value 768 MeV for $m_{\rho}$. This yields  radii
of $0.29\pm 0.02$  and $0.58\pm 0.01$ fm for the scalar and tensor
glueballs, respectively. Our results show that the scalar glueball
has a radius roughly equal to that of a pion \cite{Velikson85},
and  the tensor glueball is about two times as large as the scalar
glueball. We thus find that glueball size comparable in those of
other hadrons. The continuum limit glueball mass results in MeV
are summarized in Table \ref{tabf}. Our glueball mass results are
in good agreement with those of previous  calculations using the
improved gauge action \cite{Morningstar99} as well as with those
obtained with the Wilson action \cite{Vaccarino99}. It is
encouraging that the results are, within $10-15\%$ errors,
independent of the lattice action. We find that our glueball size
estimates are consistent with the results reported in Ref.
\cite{Ishii02} obtained by using a Gaussian extension as the
characteristic size of the glueball. We believe that our physical
glueball sizes are more accurate than those obtained in Ref.
\cite{Forcrand92} where it was found that the radius of the tensor
glueball is three times that of a pion and four times that of a
scalar glueball. The predicted zero lattice spacing results are in
fact not  found by extrapolation to zero lattice spacing, but are
instead obtained from calculations with $\beta = 2.2$ for the
glueball effective radius. Assigning a physical value to the
lattice spacing $a$ by setting the string tension to 420 MeV, the
physical radius is calculated with zero uncertainty, with no
accurate representation of the effect of the absence of
extrapolation.

\section{Conclusion}
\label{sec4}

To conclude, we have calculated the {\it SU(3)} glueball wave
functions using improved gauge and fermion lattice actions.
Instead of fixing a gauge or a path for gluons, the correlation
functions were computed from gauge invariant closed-loop paths.
This approach has an advantage over the method used in Ref
\cite{Forcrand92}, which can involve a mixing of the glueball
states with torelons, and detangling of the glueball and torelon
was done by hand. The iterative smearing procedure discussed here
can be used to obtain a rough estimate of the physical glueball
size in terms of a Gaussian extension. Although easy to implement,
it does not give detailed information about the glueball wave
function. A better approach may be to construct an elaborate
smearing procedure that maps onto the wave function in one step
(i.e., a variational method combined with fuzzing techniques).

Finally, we note that our results at the smallest lattice spacings
seem to scale. This allows for very accurate extrapolations to the
continuum limit. It is found that the physical  size of the tensor
glueball is about two times as large as that of the scalar
glueball. This is consistent with the large finite-size effect for
the tensor ($E^{++}$) glueball mass found in previous lattice
calculations. A great deal of care should be taken in making
direct comparisons with experiment, since these values ignore the
effects of light quarks and mixings with nearby conventional
mesons. We intend to include the effect of  dynamical quarks in a
future study.

\section{Acknowledgements}
We thank D. Leinweber and C. Hamer for
valuable conversations and suggestions. We are also grateful for
access to the computing facilities of the Australian Centre for
Advanced Computing and Communications (ac3) and the Australian
Partnership for Advanced Computing (APAC). This work was supported
by the Guangdong Provincial Ministry of Education.

\end{document}